# PRESENT PERFORMANCE OF ELECTRON COOLING AT COSY-JÜLICH

H. J. Stein, D. Prasuhn, H. Stockhorst, J. Dietrich, K. Fan, V. Kamerdjiev, R. Maier
Forschungszentrum Juelich GmbH, Germany

I.N. Meshkov, A. Sidorin
JINR Dubna, Russia

V.V. Parkhomchuk
BINP SB RAS Novosibirsk, Russia

Abstract

At COSY, electron cooling is used after stripping injection of $H^-$ or $D^-$ ions in order to prepare phase-space-dense ion beams before acceleration to a requested energy. The electron-cooled beam has been successfully applied for specific external experiments. The achievable beam intensity is limited by instabilities during the cooling process. Besides initial losses after injection, as long as the beam has still large emittances, the self-excitation of coherent betatron oscillations is the dominating beam loss mechanism. Perspectives for possible improvements are briefly addressed.

## 1 INTRODUCTION

The COSY synchotron accelerator and storage ring provides unpolarized and polarized proton or deuteron beams for internal or external hadron physics experiments in the momentum range from 300 MeV/c to 3.7 GeV/c [1]. Electron cooling is applied at low energies, preferably at injection energy, to prepare low-emittance coasting beams to be used after acceleration and extraction for the external experiments BIG KARL, TOF, and JESSICA, see Fig. 1. Stochastic cooling, covering the momentum range from 1.5 GeV/c up to the maximum momentum, is used to compensate energy loss and emittance growth at internal experiments as, e.g., COSY 11 [2].

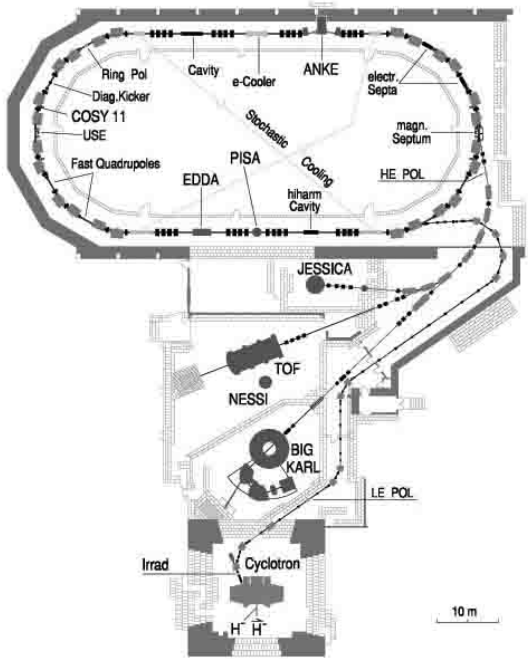

FIG. 1. Overview of the COSY accelerator facilities. At present, four internal and four external installations, where various specific experiments are being performed, require beams in a wide range of different specifications.

The electron cooler, designed and built together with the accelerator during the years 1988 until 1993 with first cooling in May 1993 [3], is now used for specific physics experiments. The first achievement has been the production of a small-diameter proton beam kicked out within one revolution by the available (relatively weak) diagnostic kicker. The resulting short proton pulses ($2 \times 10^9$ protons in 200 ns) are used by JESSICA for spallation experiments [2]. Since the end of 2001, the electron-cooled proton beam has been made available via the slow stochastic extraction for *external* experiments at the BIG KARL magnetic spectrometer, see Fig. 2, and the TOF time of flight spectrometer.

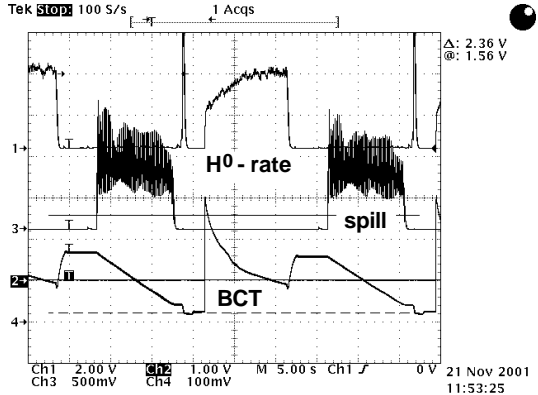

FIG. 2. Complete machine cycle to extract a 1.57 GeV/c electron-cooled proton beam to BIG KARL. The proton beam current (BCT signal) and the 10 s long spill are shown. 10 s of electron cooling after injection is made visible by the neutral particle signal ($H^0$ rate). The experimenters report 60 times more usable beam intensity due to smaller beam dimensions and - more important - much less halo compared with the uncooled beam [4].

The merits of the electron cooler for *internal* experiments, where duty cycle aspects are not as critical as in external applications, is the possibility to increase the ion beam intensity by a cooling-stacking process [2]. This procedure can be helpful in the cases of low-intensity ion sources or low-acceptance devices as storage cell targets.

## 2 THE ELECTRON COOLER

The design of the COSY electron cooler represents the state-of-the-art in the eighties [3], see Table I. The capability to produce a 3 A, 100 keV electron beam was demonstrated during various tests of the electron cooler. At present, only 25 keV beam energy is necessary for the proton injection energy of 45 MeV. Electron beam currents in the range from 50 to 440 mA were used for cooling tests. Higher electron currents are not useful because the advantage of shorter cooling times is foiled by drastically increasing proton beam losses. Currents of 170 to 250 mA have turned out to be appropriate for the physics experiments. The typical cooling time of about 10 s can be tolerated in view of the duty cycle, Fig. 2.

Important diagnostic tools to adjust and characterize the cooling process are the beam current transformer (BCT), two x-y beam position monitors in the drift solenoid, a FFT vector analyzer with integrated storage capacity as a versatile instrument to analyze and record the time evolution of longitudinal or transverse Schottky spectra, see Fig. 5, and a neutral particle ($H^0$) detector placed 24 m downstream of the electron cooler. Total $H^0$ rates and $H^0$ beam profiles in both planes are measured. The profiles represent the divergence of the ion beam at the electron cooler. Based on beta function values, emittances of the cooled beam can be determined.

Cooling was optimized by using the electron beam steering coils. Best overlap of the electron beam with the proton beam is given by the smallest coasting beam revolution frequency (minimum of the space charge depression of the electron beam). Adjusting a minimum width of the $H^0$ profiles was used as indicator for good alignment [5]. Measurements of the longitudinal cooling force by the voltage-step method in the relative velocity range $v_{rel}$ from $10^4$ to $10^6$ m/s gave, normalized to an electron density of $10^{14}/m^3$, a maximum of about $3 \times 10^{-1}$ eV/m at $v_{rel} \approx 3 \times 10^4$ m/s. These values are in reasonable agreement with results at other laboratories and what one would expect for standard nonmagnetized cooling. The initial beam momentum spread after injection of $\Delta p/p = 2 \times 10^{-3}$ is shrinking to about $10^{-4}$, revealing the intensity dependent double-peak structure of a phase-space-dense ion beam. Cooled emittances as low as 0.3 µm ($2\sigma$ value) have been obtained for beam intensities below $5 \times 10^9$ protons.

| COSY electron cooler | design parameters | used up to now | |
|---|---|---|---|
| mechanical lenght of the drift solenoid | 2.00 | | m |
| effective cooling length | ≈ 1.5 | | m |
| beam tube diameter throughout the electron cooler | 0.15 | | m |
| potential tube diameter in toroids | 0.065 | | m |
| electron beam diameter | 0.0254 | | m |
| electron beam radius in toroids | 0.60 | | m |
| magnetic field range | 80 ... 165 | 80 | mT |
| maximum electron energy | 100 | 24.5 | keV |
| gun perveance | 0.84 | | µP |
| design electron beam current at 100 keV | 4 | | A |
| design electron beam current at 25 keV | 1.8 | 0.05 ... 0.5 | A |
| collector loss factor | $\leq 5 \times 10^{-4}$ | $1 ... 4 \times 10^{-4}$ | |
| vacuum pressure in the cooling region | $5 ... 10 \times 10^{-9}$ | $5 \times 10^{-9}$ | hP |
| COSY ring | | | |
| particles | protons and deuterons (unpolarized and polarized) | | |
| type of injection | H⁻, D⁻ stripping injection, 20 ...25 µg/cm² carbon foil | | |
| injection energy | 45 MeV for protons, 76 MeV for deuterons | | |
| shape of the ring | racetrack type, two straight sections and two arcs | | |
| nominal circumference | 183.473 m | | |
| dimensions of the beam tube | round in straight sections, $d = 0.15$ m; rectangular in arcs, 0.15 m horizontal ($x$), 0.06 m vertical ($y$) | | |
| working point range | variable between 3.55 and 3.7 in both planes | | |
| optical functions at the electron cooler | $\beta_x = 8$ m, $\beta_y = 16$ m, $D = -6$ m | | |

Table I. Relevant electron cooler and COSY ring parameters

## 3 ION BEAM LOSSES AND INSTABILITIES

In this paper we concentrate on electron cooling of protons after a single injection in view of external experiments. Here, some of the operational features of the stripping injection into COSY have to be considered, see Fig. 3. The stripper foil is located behind a dipole in the extraction arc, see Fig. 1, about 40 mm off the nominal orbit. For injection the COSY orbit is bumped to the edge of the foil so that it meets the incoming cyclotron beam position and direction. The injection is controlled by three main parameters, the macropulse length $t_{macro}$, the bumper ramp down time $t_{ramp}$, and the micropulsing factor $f_{micro}$. Controlled by a shutter at the cyclotron, H⁻ (or D⁻) ions are delivered within a time interval $t_{macro}$. Simultaneously the orbit bumpers are de-energized in the same time, $t_{ramp} = t_{macro}$. If requested, the cyclotron current $I_{cycl}$ can be decreased by micropulsing, $f_{micro} = 1$ corresponds to 100% $I_{cycl}$. As injection proceeds, the betatron amplitude of the stored beam increases up to a value determined by the available horizontal acceptance. Multiscattering due to many repeated traversals through the foil and a possible mismatch of incoming and circulating beam angles with subsequent filamentation broaden the stored beam also vertically up to the available acceptance. With the standard values $t_{macro} = t_{ramp} = 20$ ms, no micropulsing, and typically 6 µA cyclotron current, the ring is filled with 5 to $10 \times 10^{10}$ protons, but at the expense of large emittances. Based on the aperture of the beam tubes, the optical functions, and the orbit distortions in COSY we estimate acceptances of $A_x = 80$ µm and $A_y = 20$ µm. The proton beam size ($3\sigma$ emittances) is then larger than the electron beam diameter. If the macropulse is made shorter at constant ramp down time one may expect a beam with smaller emittance but also less stored beam intensity.

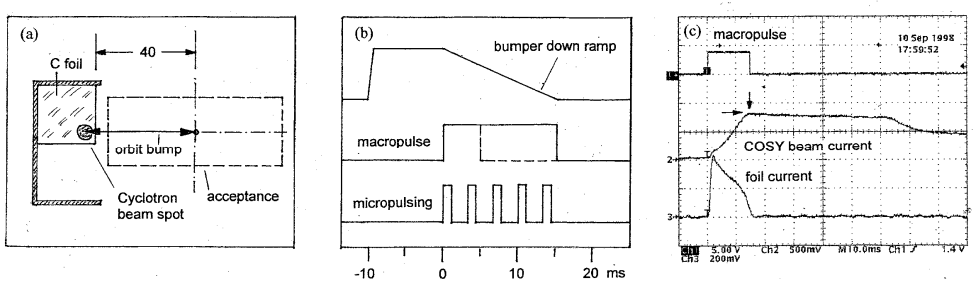

FIG. 3. Principle of the stripping injection at COSY. H⁻ or D⁻ delivered by the cyclotron injector change their charge state in a 20 μg/cm² carbon foil. Before injection the COSY orbit is bumped to the edge of the stripper foil (a). During the injection time, defined by the macropulse length, the orbit is moving back to its nominal position, coasting beam injection. Bumper ramp down time $t_{ramp}$ and macropulse length $t_{macro}$ are variable parameters (b). In (c) is shown an example for proton injection. With 6.7 μA current delivered by the cyclotron, the ring is filled in 15 ms with 8 mA circulating beam ($\approx 10^{11}$ protons at 45 MeV). Micropulsing by chopping the macropulse allows to reduce the intensity $I_{cycl}$ of the incoming cyclotron beam.

Bunching the high-intensity uncooled coasting beam to accelerate it, regularly leads to immediate losses of about 50% caused by the increased momentum spread and dispersion. When electron cooling is applied here, strong beam losses are also observed, however, occurring slowly and extended over the whole cooling time, see Fig. 4(a). When then the beam is accelerated the bunching losses should be much less or even zero due to the smaller momentum spread. In the example shown in Fig. 4(a) with 7 mA injected current, the bunching efficiency is poor. In addition, a kink in the BCT curve as well as in the H⁰ rate is observable 7 s after injection. This phenomenon was identified as the onset of a strong vertical coherent betatron oscillation, see Fig 5(a). In parallel, a horizontal betatron oscillation is detected already after 2 s. These oscillations, which do not stop until the end of the cooling time, obviously are the cause for the poor bunching efficiency in this case. The obtained beam current after acceleration to 1.64 GeV/c was 3 mA.

Figure 4(b) shows the remarkable influence of a change of the injection parameters. When the macropulse was reduced to 6 ms (in addition some micropulsing was applied) a quite different, more regular BCT signal was obtained close to that what one would expect for a "smooth" cooling process: Injected current as low as 1.2 mA, only slight losses right after injection, then a flat BCT signal with a steadily increasing H⁰ rate, only a short horizontal oscillation after ≈ 3 s, see Fig. 5(b), bunching perfect without any loss, final emittance in both planes $\varepsilon_{x,2\sigma} = \varepsilon_{y,2\sigma} = 0.5$ μm after 10 s cooling, all that resulting in only 35% less proton current after acceleration compared to the case with very high injected current.

In both cases losses are already observed just after injection before any betatron oscillations are visible. By this time the proton beam has still large emittances. Here, the electron beam can be considered as a beam disturbing "target" causing particle losses similar to those which are observed when the electron energy is detuned, a phenomenon known as "electron heating" [6]. These initial losses increase substantially at higher electron currents. This is the reason why useful electron currents are lower than one would like to apply in view of short cooling times. In this context we refer to Ref. 7. For the proton case in COSY we conclude that under the given operational conditions not much more than $1 \times 10^{10}$ protons can be reasonably well cooled. 170 to 200 mA electron beam and $2 \times 10^{10}$ injected protons turned out as maximal values to get $1.5 \times 10^{10}$ protons into the flat top. More electron current increases the initial losses, too many protons lead to the observed (coherent) beam oscillations. In a recent beam experiment it could be shown that sextupoles can suppress the most dangerous vertical oscillations and, therefore, increase the instability threshold.

Further studies on instability effects *after* the beam is well cooled have shown that transverse oscillations are always excited when the proton number is too large [8]. Even at beam intensities lower than $10^9$ protons, transverse oscillations may suddenly be excited still minutes after a stable equilibrium is reached. However, for the practice in COSY these "delayed" instabilities are not relevant because the beam is accelerated as soon as the ion beam is cool enough.

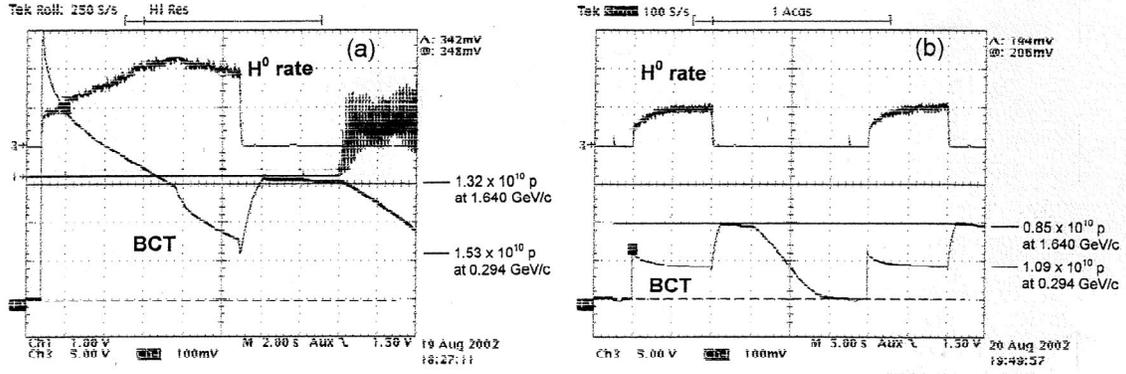

FIG. 4. Case study for electron cooling during a recent BIG KARL experiment ($I_e$ = 170 mA). High and low intensity injections controlled by the injection parameters. (a) long macropulse $t_{macro}$ = 20 ms, no micropulsing, cyclotron current ≈ 6 µA. (b) short macropulse $t_{macro}$ 6 ms and 20% reduction of the cyclotron current by micropulsing, $f_{micro}$ = 0.8. Bumper ramp down time $t_{ramp}$ = 20 ms in both cases. BCT signal: 100 mV/division = 1 mA proton current = 1.27 x $10^{10}$ particles at injection, $p_{inj}$ = 294 MeV/c. $H^0$ rate: 10 V = $10^4$ particles. The beam intensities after acceleration to 1.64 GeV/c are not much different due to large particle losses in case (a).

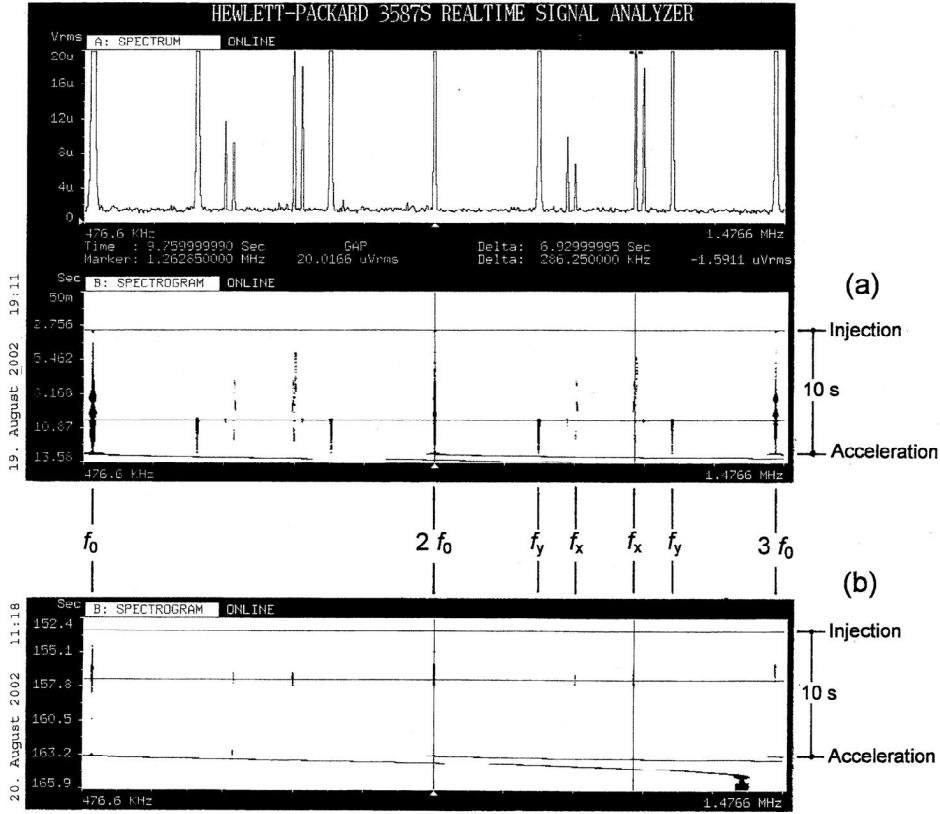

Fig. 5. FFT spectrograms of the transverse Schottky noise recorded during the 10 s cooling time for (a) high and (b) low intensity injection. The frequency span comprises the first three harmonics $h$ of the revolution frequency $f_0$ = 488.3 kHz. The longitudinal signals appear because the COSY orbit is displaced at the Schottky detector. The frequencies between $h$ = 1 and $h$ = 2, and $h$ = 2 and $h$ = 3 are the betatron frequencies $f_x$ and $f_y$ of the horizontal and the vertical tune, $Q_x$ = 3.59, $Q_y$ = 3.69. The onset of the vertical oscillation (a) corresponds to the kink of the BCT curve in Fig. 4(a).

## 4 CONCLUSIONS AND OUTLOOK

Small-emittance electron-cooled proton beams have opened new experimental possibilities at the external experiments The beam current finally available at the targets is predominantly determined by the number of protons which survive the cooling process. 1 to 1.5 x $10^{10}$ protons in $2\sigma$ emittances of 0.5 μm seem to be the limit under the present operational conditions at COSY.

As a test for upcoming tasks, electron cooling of deuterons was tried for the first time in January 2002. As a result of the higher injection energy of 76 MeV, the injected and cooled intensity was appreciably higher. 1.6 x $10^{11}$ deuterons could be injected and were cooled with 190 mA electron current at 20.6 keV electron energy. The losses and the visible instability jumps during cooling were clearly weaker. 5.5 x $10^{10}$ particles at the end of 14 s cooling time were lossless bunched and accelerated with 90 % efficiency [1]. However, the emittances after cooling were appreciably larger and not equal, $\varepsilon_{x,2\sigma} \approx 2$ μm, $\varepsilon_{y,2\sigma} \approx 5$ μm.

If higher proton intensities should be needed, the causes for the proton losses have to be investigated in more detail. Are the large initial emittances at present unavoidably to obtain high intensity proton beams or is the phase space density alone the limiting factor for the cooled beam intensity? What will be the merit of a new injector with higher injection current? Studies on elementary measures like orbit and tune control, optimization of the injection, and perhaps modification of the beta functions at the cooler section are necessary to make COSY effective for the new LINAC injector [9]. In addition, detailed studies on "impedances" are necessary since the COSY ring is filled up with experiment chambers and other insertions which disturb the homogeneity of the beam tube. Besides the possibility of "heating" the ion beam (e.g., increasing $\Delta p/p$ by electronic noise), a feedback kicker system is a well-proven technique [10] to counteract the coherent betatron oscillations. A feedback kicker is already constructed and is waiting for experimental studies.